\documentclass[a4paper,11pt]{article}
\usepackage{pos}
\usepackage{braket}

\title{Finite volume effects near the chiral crossover }

\author*[a]{Ruben Kara}
\author[a]{Szabolcs Borsányi}
\author[a,b,c,d]{Zoltán Fodor}
\author[a]{Jana N. Guenther}
\author[b,e]{Paolo Parotto}
\author[d]{Attila Pásztor}
\author[a]{C. H. Wong}

\affiliation[a]{University of Wuppertal, Department of Physics, Wuppertal D-42119, Germany}
\affiliation[b]{Pennsylvania State University, Department of Physics, State College, PA 16801, USA}
\affiliation[c]{Jülich Supercomputing Centre, Forschungszentrum Jülich, Jülich D-542425, Germany}
\affiliation[d]{E{\"o}tv{\"o}s University, Budapest 1117, Hungary}
\affiliation[e]{Dipartimento di Fisica, Universit\`a di Torino and INFN Torino, Via P. Giuria 1, I-10125 Torino, Italy}

\emailAdd{rkara@uni-wuppertal.de}

\abstract{The effect of a finite volume presents itself both in heavy ion experiments as well as in recent model calculations. The magnitude is sensitive to the proximity of a nearby critical point. We calculate the finite volume effects at finite temperature in continuum QCD using lattice simulations. We focus on the vicinity of the chiral crossover. We investigate the impact of finite volumes at zero and small chemical potentials on the QCD transition through the chiral observables.}

\FullConference{The 40th International Symposium on Lattice Field Theory (Lattice 2023)\\
July 31st - August 4th, 2023\\
Fermi National Accelerator Laboratory\\}

\begin{document}
\maketitle

\section{Introduction}
\noindent
It is well known that QCD exhibits a thermal transition which turned out to be an analytic crossover in the case of physical quark masses and vanishing baryon chemical potential $\mu_B=0$ \cite{Aoki:2006we}. Finite size scaling using aspect ratios $LT=4,5,6$ specified the transition as analytic since the peak of the chiral susceptibility shows basically no or a mild volume dependence. Further studies of the equation of state demonstrate that the main driver of uncertainties are not finite volume effects, but instead cut-off effects which lead to taste-violation in the case of staggered quarks \cite{Borsanyi:2010cj}. These can be significantly reduced by employing tree-level corrections, stout-smearing methods or using the HISQ action \cite{Borsanyi:2013bia,HotQCD:2014kol}. Especially the observation that there is basically no volume dependence in the transition region contributed to the unspoken common standard in the community to choose $LT=4$ to study the thermal properties of QCD such as high-order fluctuations \citep{Borsanyi:2018grb,Bazavov:2020bjn} or the pseudocritical transition line $T_c(\mu_B)$ \cite{Borsanyi:2020fev,HotQCD:2018pds,bonati}.\\
Nevertheless finite volume effects play a crucial role phenomenologically and theoretically. The fireball produced in heavy-ion collisions is of finite size and if the crossover turns into a real transition, volume effects get more and more severe. Hence we study the impact of finite volumes at vanishing chemical potential and at finite $\mu_B$ using the imaginary chemical potential Taylor method by setting the focus on the chiral observables.



\section{Chiral observables}
\noindent
In the case of vanishing quark masses the chiral condensate $\braket{\bar{\psi}\psi}$ deals as a true order parameter to probe the spontaneous breaking of the underlying chiral symmetry. Since nature presents us small but finite quark masses, the symmetry is also explicitly broken which leads to a non-vanishing value of the condensate at high temperatures $T$ although the spontaneous breaking is restored. We are interested in physical results and perform whenever it is possible a continuum extrapolation. Hence we use the following renormalization scheme to remove additive and multiplicative divergences
\begin{align}
\braket{\bar{\psi}\psi}&=\frac{T}{V} \frac{\partial \log Z}{\partial m} \qquad \qquad \braket{\bar{\psi}\psi}_R=-\left[\braket{\bar{\psi}\psi}_T -\braket{\bar{\psi}\psi}_{T=0} \right] \frac{m}{f_\pi^4}\\
\chi&=\frac{T}{V} \frac{\partial^2 \log Z}{\partial m^2} \qquad \qquad \;\;\;\;\;\chi_R=\left[\chi_T -\chi_{T=0} \right] \frac{m^2}{f_\pi^4},
\end{align}
\noindent
by subtracting the zero temperature part of the observable $\braket{...}_{T=0}$ and multiplying with the light quark mass $m$ in lattice units. To get a dimensionless quantity, the result is divided by the pion decay constant $f_\pi$.

\section{Volume dependence of the chiral condensate}
\noindent
The key feature of a crossover transition is basically no or a very mild volume dependence of the observable and hence the absence of discontinuities or divergences up to the infinite volume limit.
In the opposite direction, i.e. decreasing the volume, the behavior is not so clear. 
Chiral perturbation theory (chiral PT) predicts an exponential dependence of the condensate as a function of the spatial extension $N_x$. The leading asymptotic behavior of the condensate at vanishing magnetic field and $T=0$ takes on the form \cite{Adhikari:2023fdl}
\begin{equation}
\braket{\bar{\psi} \psi} \sim \frac{\sqrt{m_\pi}}{F_\pi^2} \frac{e^{ -m_\pi N_x }}{\left( 2\pi N_x \right)^{3/2}}.
\label{eq:chiral_pt}
\end{equation}
\noindent
This can be compared with our lattice results if we pick a temperature below $T_c$ as shown in Fig. \ref{fig:v_scaling_pbp_t140}. Here the chiral condensate is solved via a spline interpolation at fixed $T=140$ MeV for all lattices with $N_t=12$.




\begin{center}
\includegraphics[width=0.6\textwidth]{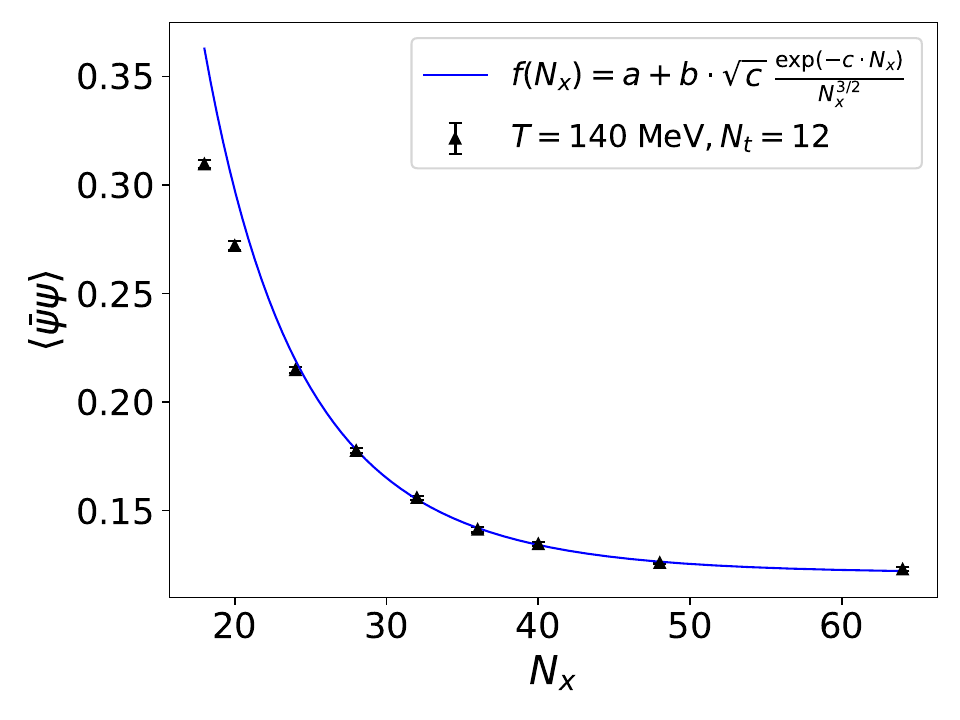}
\captionof{figure}{Chiral condensate solved at a fixed temperature $T=140$ MeV for every lattice with $N_t=12$ as a function of the spatial extension $N_x$. The blue curve is a fit inspired by chiral PT (Eq. (\ref{eq:chiral_pt})) in the range of $N_x \in [28,64]$.}
\label{fig:v_scaling_pbp_t140}
\end{center}
\noindent
The blue curve is the fit function $f(N_x)$ as shown in the legend and provides $\chi^2/\mathrm{ndof}=1.03$. The coefficient $c$ is $m_\pi$ according to Eq. (\ref{eq:chiral_pt}) and reads $c=131 \pm 10 \mathrm{\;MeV}$. This remarkable agreement with the pion mass is only true for $N_x \geq 28$. One reason for this lies in the fact that the transition temperature for $18^3 \times 12$ and $20^3 \times 12$ is below $T=140$ MeV as shown in Fig. \ref{fig:obs_nx} in the lower left panel. Hence the system tends to be deconfined and cannot be described by the chiral PT Eq. (\ref{eq:chiral_pt}) which is valid for zero temperature.\\ 
The exponential behavior of the condensate as a function of $N_x$ could be observed for a large range of temperatures. Hence it is reasonable to fit $g(N_x)=a'+b' \cdot \exp(-c'\cdot N_x)$ inspired by Eq. (\ref{eq:chiral_pt}) to the condensate values at fixed temperature. The results for the coefficients $c'$ and $b'$ are shown in Fig. \ref{fig:bc_coff}:
\begin{center}
\includegraphics[width=0.51\textwidth]{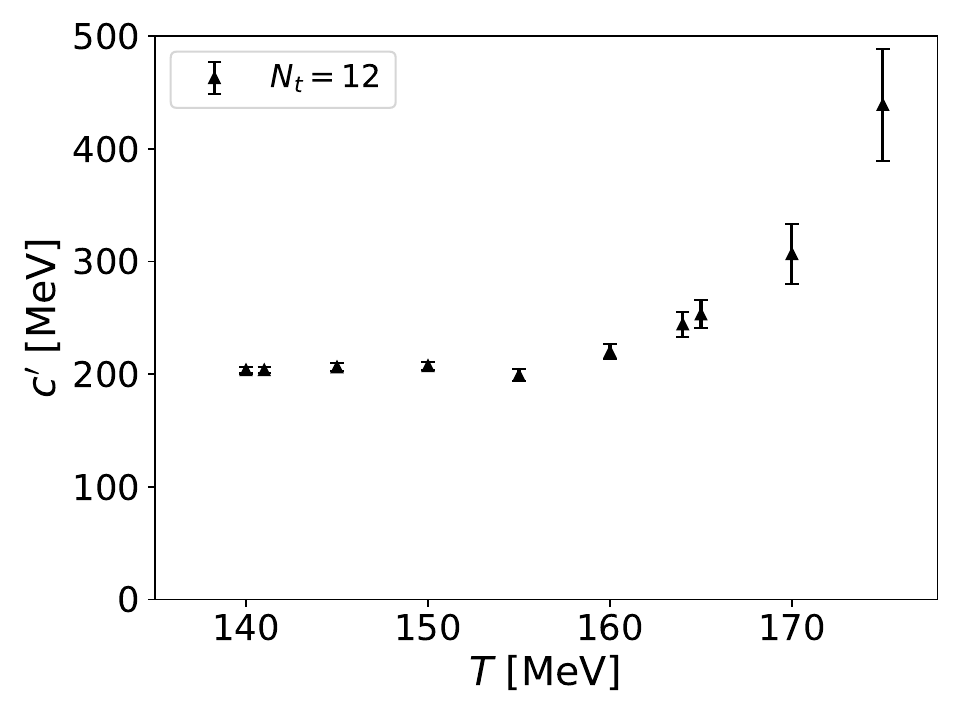}\includegraphics[width=0.51\textwidth]{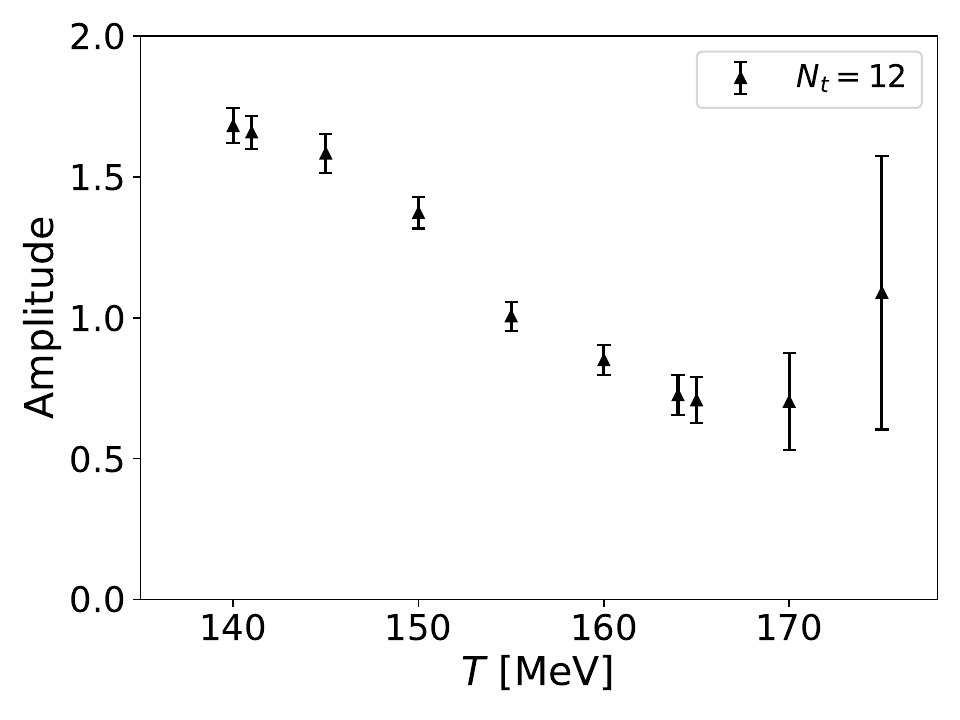}
\captionof{figure}{Fit parameters of $g(N_x)=a'+b' \cdot \exp(-c'\cdot N_x)$ as functions of the temperature. Left: $c'$ coefficient converted in MeV. Right: Corresponding amplitude $b'$.}
\label{fig:bc_coff}
\end{center}
\noindent
The fit exponent $c'$ is nearly constant and takes on a value of around the QCD scale $200$ MeV and shows a rapid rise after passing $T_c \approx 156$ MeV. In the case of the amplitude $b'>1$ for $T<T_c$. If the temperature is higher than $T_c$, then $b'$ shrinks to values below $1$.

\section{Volume dependence of the transition temperature $T_c$}
\noindent
To obtain the transition temperature $T_c$ we follow a similar strategy as described in \cite{Borsanyi:2020fev}. The chiral susceptibilty is expressed as a function of the condensate. The advantage is that $\chi(\braket{\bar{\psi}\psi})$ has a simpler form compared to $\chi(T)$ and can be fitted more precisely. Together with the corresponding $\braket{\bar{\psi}\psi}_c$ for which $\chi$ takes on its maximum value, the transition temperature can be read off from $\braket{\bar{\psi}\psi}(T)$  via spline interpolation. This procedure allows us to calculate precisely the proxy $\delta T$ for the width of the transition defined as
\begin{align}
\delta T&= \braket{\bar{\psi}\psi}^{-1}\left(\braket{\bar{\psi}\psi}_c +\frac{\Delta\braket{\bar{\psi}\psi}}{2} \right)-\braket{\bar{\psi}\psi}^{-1} \left(\braket{\bar{\psi}\psi}_c - \frac{\Delta\braket{\bar{\psi}\psi}}{2} \right)
\label{eq:delT}\\
\Delta\braket{\bar{\psi}\psi}&= \sqrt{-\chi_\mathrm{max} \left(\left.\frac{\mathrm{d^2}\chi}{\mathrm{d}\braket{\bar{\psi}\psi}^2}\right\vert_{\braket{\bar{\psi}\psi}_{c}} \right)^{-1}}.
\end{align}
\noindent
More details can be found in \cite{Borsanyi:2020fev,kara,jana}. For a broad range of aspect ratios we can now perform a continuum extrapolation as examplary demonstrated in the left panel of Fig. \ref{fig:tc_conti}. The continuum extrapolotated results of the transition temperature for each aspect ratio are shown on the right panel. Again we observe an exponential dependence which allows us to obtain the infinite volume limit of the continuum extrapolated transition temperatures
\begin{equation}
T_c(N_t \rightarrow \infty, LT \rightarrow \infty ) = 158.9 \pm 0.6\;\mathrm{MeV}.
\end{equation}

\begin{center}
\includegraphics[width=0.51\textwidth]{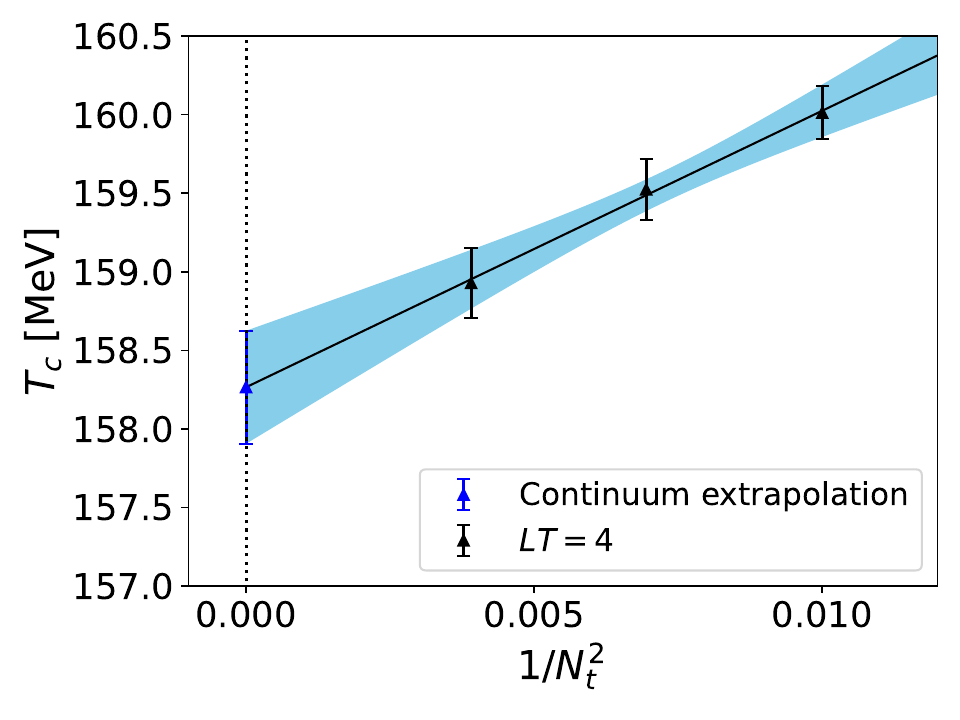}\includegraphics[width=0.51\textwidth]{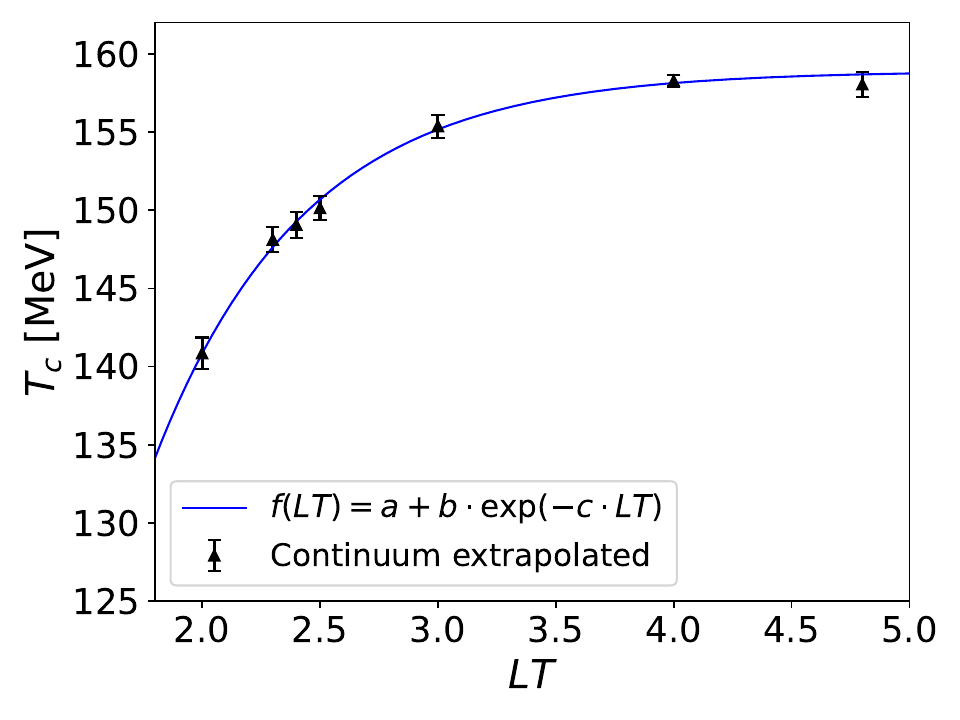}
\captionof{figure}{Left: Exemplary continuum extrapolation of $T_c$ at aspect ratio $LT=4$. Right: Continuum extrapolated $T_c$ as a function of the aspect ratio $LT$ and additional infinite volume extrapolation via an exponential fit.}
\label{fig:tc_conti}
\end{center}
\noindent
The exponential dependence on the volume is not limited to $T_c$. As demonstrated in Fig.\ref{fig:obs_nx}, the peak of the susceptibility $\chi_\mathrm{max}$ and the width of the transition $\delta T$ Eq. (\ref{eq:delT}) indicate a similar behavior. 
\begin{center}
\includegraphics[width=0.5\textwidth]{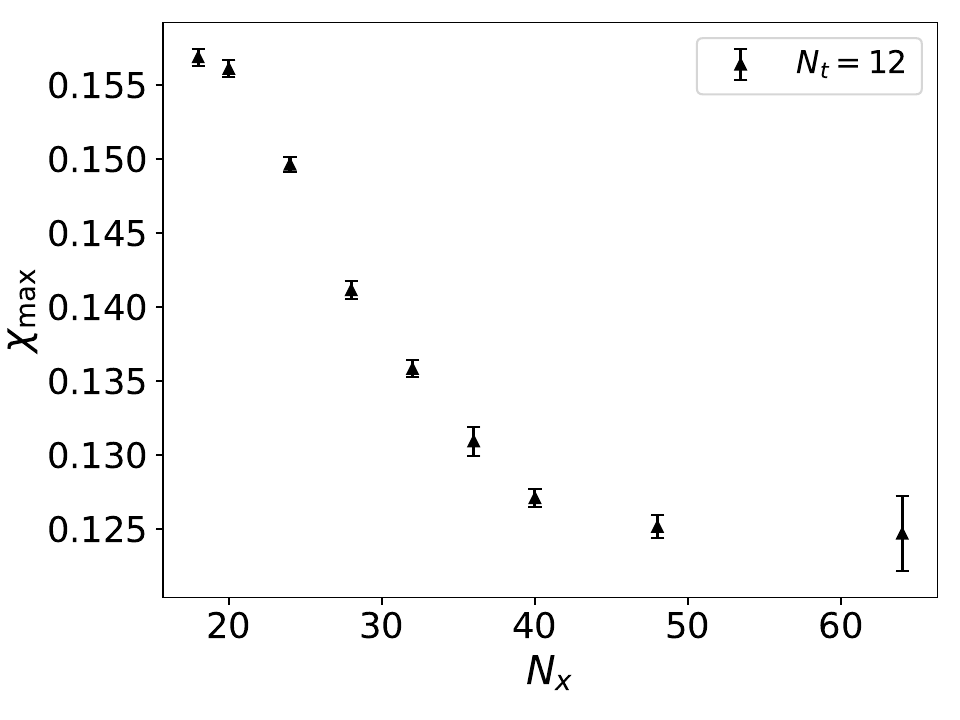}\includegraphics[width=0.5\textwidth]{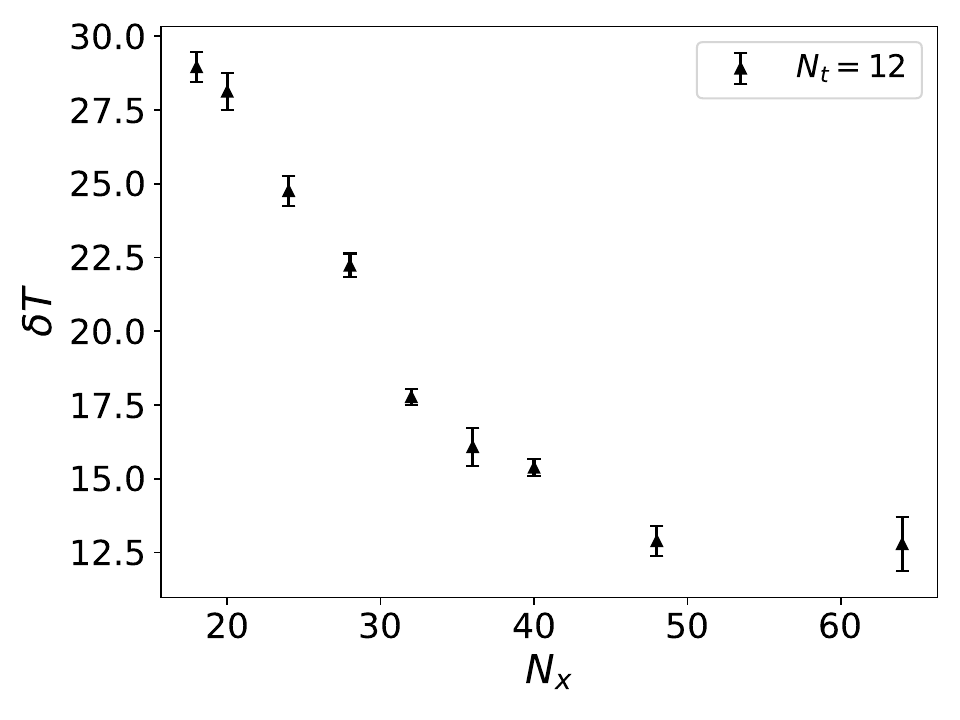} 
\includegraphics[width=0.5\textwidth]{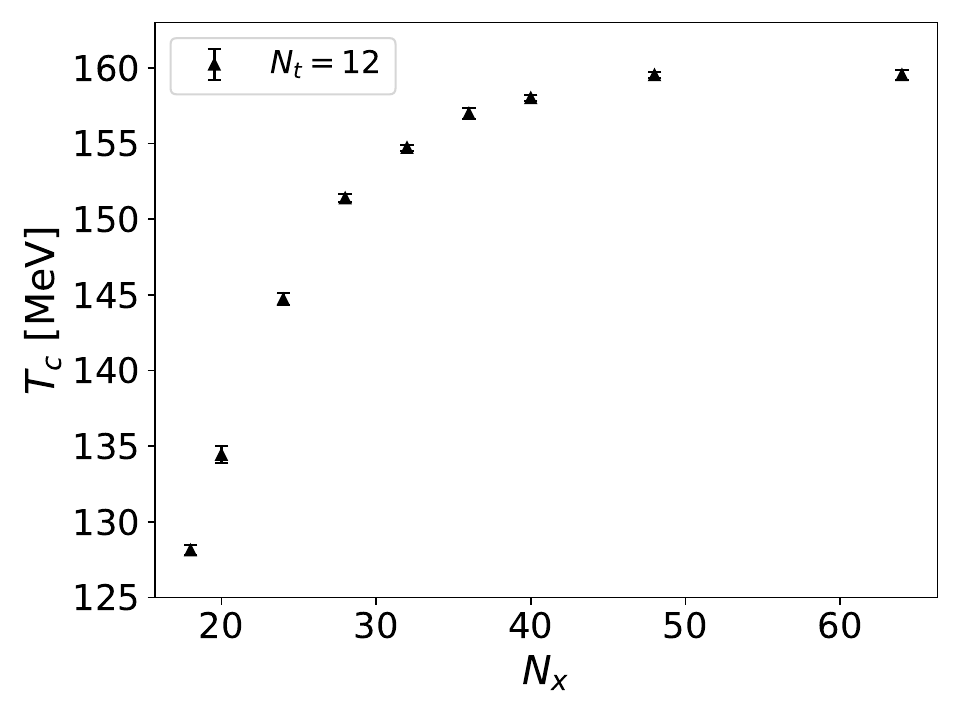}\includegraphics[width=0.5\textwidth]{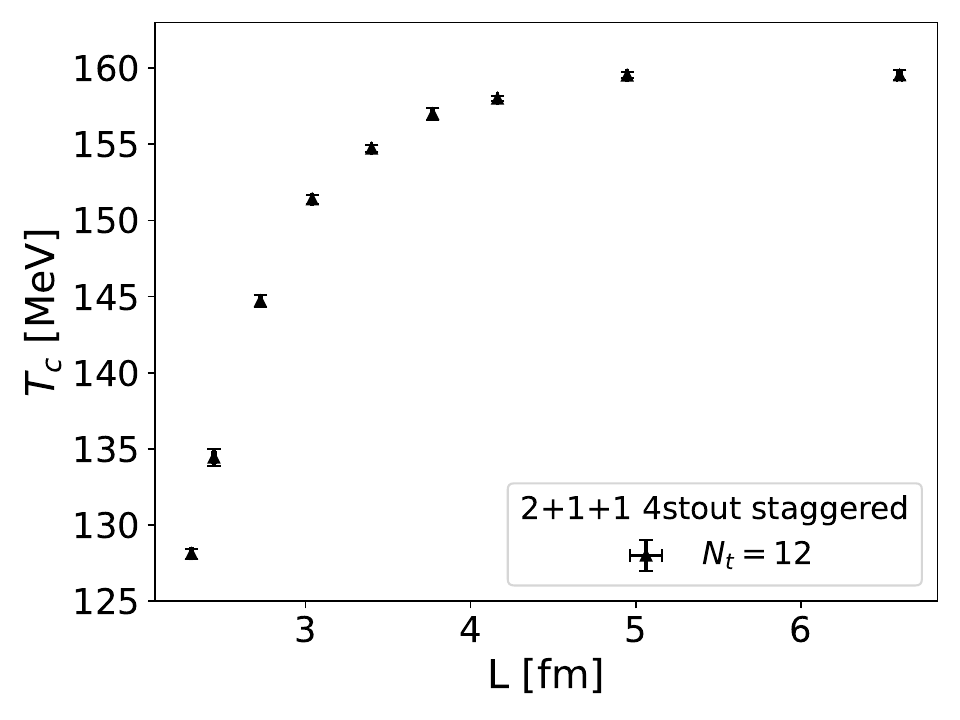}
\captionof{figure}{Volume dependence of $\chi_\mathrm{max}$ (upper left), $\delta T$ (upper right) and $T_c$ (lower left) as functions of $N_x$. The lattice geometry is converted in the box size L in fm (lower right).}
\label{fig:obs_nx}
\end{center}
\noindent
The peak of the susceptibility (upper left panel) decreases and stays nearly constant if $N_x \gtrapprox 40$ ($LT \gtrapprox 3.3 $) which is a clear sign of a crossover. It confirms the common standard to use $LT=4$ in QCD thermodynamics to be close to the infinite volume limit. In the opposite direction, the peak increases significantly as the volume is further decreased.

\section{Volume dependence of $T_c$ at finite and real $\hat{\mu}_B$}
\noindent
So far we set the focus on vanishing chemical potential. Let us extend our results to finite density and investigate the volume dependence. To circumvent the sign problem, we performed simulations at purely imaginary and vanishing chemical potentials. These runs deal as a lever arm to extrapolate to finite and real $\hat{\mu}_B$. In Fig. \ref{fig:tc_mu_sq} the temperature is shown as a function of $\hat{\mu}_B^2$ for various volumes at fixed $N_t=12$. There is a clear volume dependence visible for imaginary and vanishing $\hat{\mu}_B$. In the linear extrapolation in $\hat{\mu}_B^2$ we see that the volume dependence gets weaker and tends to disappear near $\hat{\mu}_B^2=6$.

\begin{center}
\centering
\includegraphics[width=0.65\textwidth]{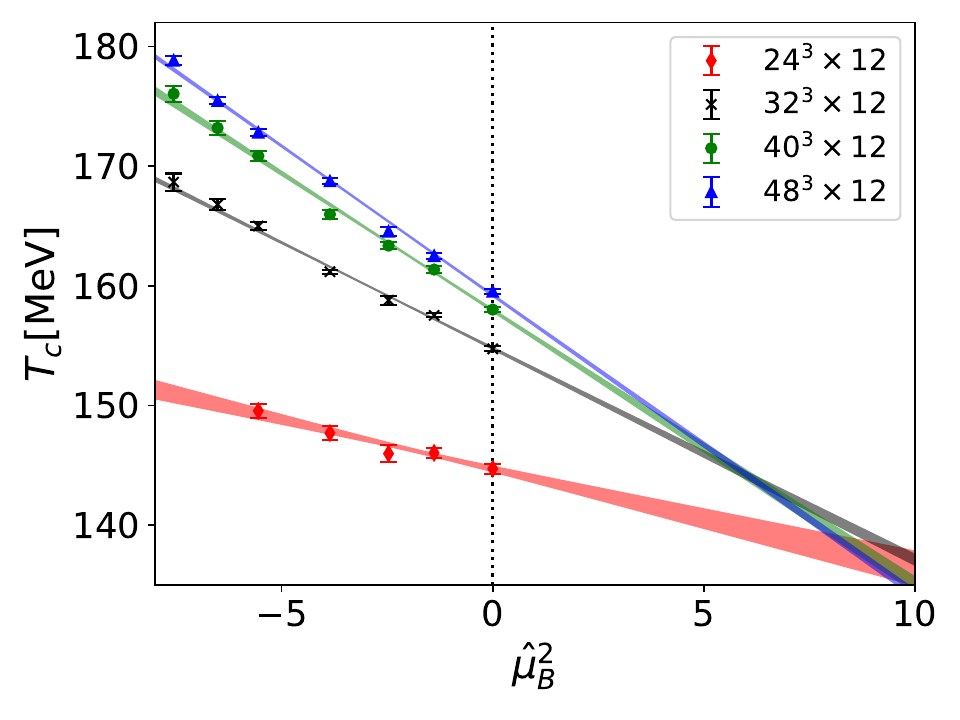}
\captionof{figure}{$T_c$ as a function of $\hat{\mu}_B^2$ for various lattices with $N_t=12$. The colored bands indicate a linear extrapolation.}
\label{fig:tc_mu_sq}
\end{center}
\noindent
Given these runs, we extrapolate $T_c$ to real $\mu_B$ according to $\frac{T_c(\mu_B)}{T_c(0)}=1-\kappa_2 \left( \frac{\mu_B}{T_c(\mu_B)}  \right)^2$ up to leading order. The results are shown in 
Fig. \ref{fig:tc_cut_vol} and \ref{fig:tc_fm}.

\begin{center}
\includegraphics[width=0.52\textwidth]{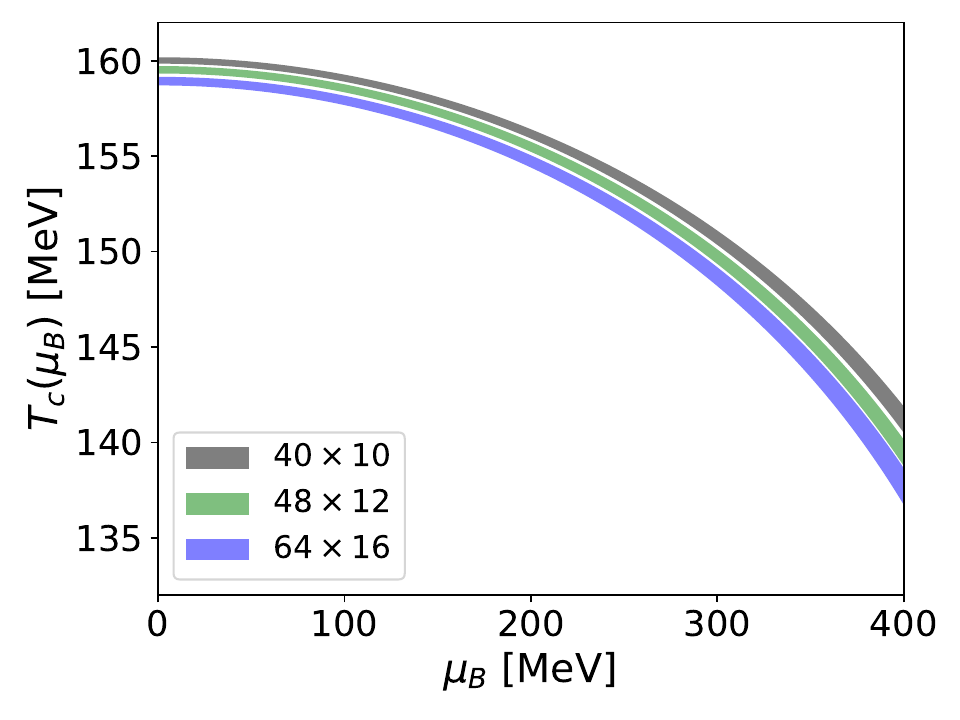}\includegraphics[width=0.52\textwidth]{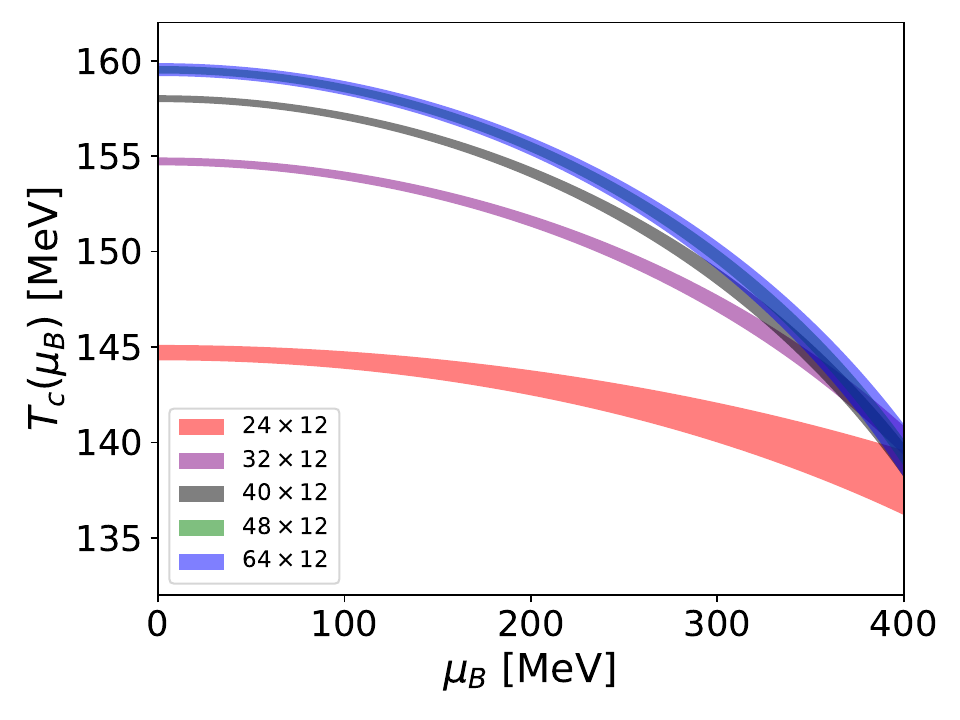}
\captionof{figure}{$T_c$ extrapolated to finite and real $\mu_B$ from imaginary chemical potentials.\\ Left: Fixed aspect ratio $LT=4$ and the discretization effects. Right: Fixed $N_t=12$ and varying volume.}
\label{fig:tc_cut_vol}
\end{center}
\noindent 
Fixing the aspect ratio $LT=4$ and varying the temporal extension $N_t$ indicates that the cut-off effects seem to stay rather constant in the extrapolation regime of $\mu_B$ as the bands of the left panel of Fig. \ref{fig:tc_cut_vol} are nearly parallel. This is not the case if the focus is set on finite volume effects for which $N_t=12$ is fixed and $N_x$ is varied as depicted on the right panel. Here the volume effects seem to decrease for higher chemical potentials and even to completely disappear around $\mu_B \approx 400$ MeV.

\begin{center}
\includegraphics[width=0.65\textwidth]{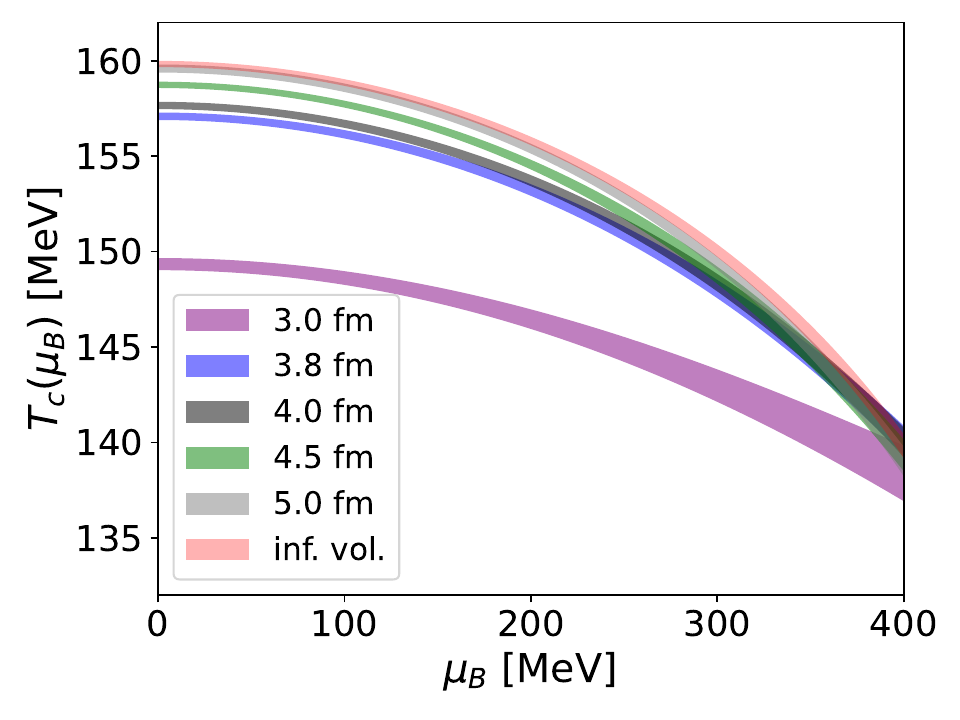}
\captionof{figure}{$T_c$ extrapolated to finite and real $\mu_B$ for various box sizes converted in fm for $N_t=12$.}
\label{fig:tc_fm}
\end{center}
\noindent
Given the transition lines for every lattice with $N_t=12$, we can calculate the result in a finite box with size $L$ [fm]. The idea is to keep the box size constant and to vary the lattice geometry at fixed temporal extension. For this purpose $T_c(N_x)$ is iterated for each $\mu_B$ to match the desired box size in fm. The results are shown in Fig. \ref{fig:tc_fm}. Here we can conclude that a box size of $L=5$ fm agrees with the infinite volume extrapolated result.

\paragraph*{Acknowledgments\\}
\noindent
This work is supported by the MKW NRW under the funding code NW21-024-A. Further funding was received
from the DFG under the Project No. 496127839. This work was also supported by the Hungarian National Research,
Development and Innovation Office, NKFIH Grant no KKP126769. This work was also supported by the NKFIH excellence grant TKP2021\_NKTA\_64. The authors gratefullly acknowledge the Gauss Centre for Supercomputing e.V. (www.gauss-centre.eu)
for funding this project by providing computing time on the GCS Supercomputer HAWK at H\"ochstleistungsrechenzeitrum Stuttgart.

\end{document}